\documentclass[english,aps,prb,twocolumn,superscriptaddress,amsmath,amssymb,showpacs,showkeys]{revtex4}
\usepackage[T1]{fontenc}
\usepackage[latin1]{inputenc}
\usepackage{graphicx}
\usepackage{subfigure}

\makeatletter

\providecommand{\tabularnewline}{\\}

\usepackage{epsfig}

\usepackage{prettyref}

\usepackage{babel}
\makeatother

\begin{document}

\title{Evidence of kinetically arrested supercooled phases in the pervoskite oxide NdNiO$_3$}

\author{Devendra Kumar}

\affiliation{Department of Physics, Indian Institute of Technology Kanpur 208016,
India}

\author{K.P. Rajeev}

\email{kpraj@iitk.ac.in}

\affiliation{Department of Physics, Indian Institute of Technology Kanpur 208016,
India}

\author{J. A. Alonso}

\affiliation{Instituto de Ciencia de Materiales de Madrid, CSIC, Cantoblanco,
E-28049 Madrid, Spain}

\author{M. J. Martínez-Lope}

\affiliation{Instituto de Ciencia de Materiales de Madrid, CSIC, Cantoblanco,
E-28049 Madrid, Spain}

\begin{abstract}
We report the time and temperature dependent response of thermopower
in the non-magnetic perovskite oxide NdNiO$_3$. We find that on
cooling below the metal-insulator transition temperature the system evolves into a phase separated
state which consists of supercooled metallic and insulating
phases. This phase separated state exhibits out of equilibrium
features such as cooling rate dependence and time dependence. The existence of these
dynamical features is attributed to the transformation of
supercooled metallic phases to the insulating state. On cooling
a small fraction of supercooled phases gets kinetically arrested in a glassy state and these supercooled phases remain in that state down to low temperature. In the heating cycle the
arrested  states dearrest above 150~K and this results in the reappearance of time dependent
features.

\end{abstract}



\maketitle

\section{INTRODUCTION}

First-order phase transition (FOPT) has been a topic of
fundamental importance in the field of condensed matter physics.
The materials undergoing such a transition often exhibit a
phase-separated state in the vicinity of the transition. The phase
separated state in many FOPT compounds displays a variety of
interesting phenomena such as colossal magnetoresistance,
CMR,\cite{Dagotto,Zhang,Uehara1} time dependent resistivity and
magnetization,\cite{Morin, Levy, Ghivelder} dependence of physical
properties on rate of temperature change and enhanced 1/$f$
noise.\cite{Fischer, Uehara, Podzorov} From the last decade
onwards, the main focus of research in FOPT studies  is centered
around understanding  the origin  of the phase separated state and
its response to various perturbations such as magnetic and
electric fields, pressure, x-ray and laser illumination, strain,
temperature variation etc.\cite{Kimara, Dagotto1, Casa,
Tomioka,Kiryukhin,Tokunaga,Dong} These studies reveal that
coexisting phases have nearly equal free energies and the
application of an external perturbation decreases the free energy
of one phase relative to the other phase giving rise to a large
change in the observed physical properties. Later it has been
found that even in the absence of any apparent perturbations, such
as mentioned above, the relative volumes of the coexisting phases
can shift such that the total free energy decreases, which results
in time dependence of physical properties at a fixed
temperature.\cite{Devendra,Levy}

The existence of a mixed phase region with nearly equal free
energies have its origin in the phenomenon of metastability, in
particular supercooling or superheating. In a first order
transition where heterogeneous nucleation is missing, the high
temperature phase can survive below the thermodynamic transition
temperature $T_C$ as a supercooled (SC) metastable
phase.\cite{Chaikin} The homogeneous nucleation will start when
the thermal energy is of the order of the free energy barrier that
separates the SC state from the ground state.\cite{Chaddah} Each
SC phase has a characteristic temperature $T^*$  known as the
limit of metastability. Below this temperature the energy barrier
separating the metastable state from the ground state vanishes and
the SC phase is free to flip to the ground state. The SC phase
will flip to the ground state only when their kinetics are not
frozen $ie$ their glass formation temperature $T_g$  is less than
the  limit of metastability $T^*$.\cite{Chaddah1} Now if the glass
forming temperature is higher than the limit of metastability, the
SC phase will remain trapped down to low temperatures even when
the free energy barrier vanishes. On subsequent heating the
kinetics of these trapped phases will be restored to the
metastable state above the temperature $T > T_g$. In a phase
separated system where the SC phases are not trapped in a
glass-like state, the thermally activated stochastic switching of
SC phases to the ground state will give rise to non-Debye
relaxation in physical properties.\cite{Devendra} So for $T_g <
T^*$ time dependent effects will be observed only in the cooling
cycle, while for $T_g > T^*$  time dependent effects will be
observable both in heating and cooling cycle.

Here we have used the compound NdNiO$_3$ to understand
supercooling and glass formation in a polycrystalline material
where a distribution of SC regions coexist, with each one of the
different SC regions having a unique $T^*$.\cite{Devendra}
NdNiO$_3$ belongs to the family of rare earth nickelates and it
undergoes a temperature driven metal to insulator transition and a
para to antiferromagnetic transition at
200~K.\cite{Medarde,Granados} Now for the study of the dynamics of
a phase separated system such as NdNiO$_3$ the ideal probe to
determine the fractional volume of the constituting phases could
perhaps be a muon spin relaxation experiment.\cite{Torrance} But
for the lack of access to such facilities we depend on transport
properties such as resistivity and thermopower to probe the
system. These measurements suffer from the  effects of percolation
and inter-crystallite contact resistance and do not allow us to do
an accurate estimate of the constituent volume fractions at
temperatures far below the percolation threshold.\cite{Hurvits}
While thermopower and resistivity measurements  on NdNiO$_3$ as
functions of temperature has
 been reported earlier by others, in this work we
 focus our attention on the dynamical aspects of the M-I phase transition
through time dependent thermopower measurements. We
find that at temperatures far below the percolation threshold,
thermopower measurements are more sensitive to the presence of
minority phases than that of resistivity.  Further investigations
suggest that on cooling the sample below the metal-insulator transition temperature, $T_{MI}$, while a majority of
SC regions transform to the insulating state, a small fraction of
SC regions get kinetically  arrested in a glassy state. These SC
regions present in the glassy metallic state exist down to low temperatures.

 \section{Experimental Details }
Polycrystalline NdNiO$_3$ samples in the form of 6 mm diameter and
2 mm thick pellets were used in the thermopower measurements. The
details of sample preparation and characterization are described
elsewhere.\cite{Massa} A temperature of 1000$^\circ$C and oxygen
pressure of 200 bar is required to get good quality samples.

Below $T_{MI}$ (200~K) NdNiO$_3$ is not in thermodynamic
equilibrium and it relaxes with time and thus any experimental
result would depend on the procedure used for the measurement. We
used the following procedure. While cooling we start with 300~K
with the two sides of the NdNiO$_3$ pellet kept at a temperature
difference of 2~K, and then record the voltage ($V_{meas}$) in
steps of 1~K after allowing the temperature to stabilize at each
point for about 30~s.  In between two temperature points the
sample was cooled at a fixed cooling rate of 2~K/min. During this
process also care was taken to ensure that the temperature
difference between the two ends of the sample remained at
2~K.\cite{endnote1} This procedure ensures that the two ends of
the sample smoothly vary with temperature without any intermediate
thermal oscillation. This condition is important since in our
earlier measurements we have observed that thermal oscillations
change the physical state of the system.\cite{Devendra} To remove
the effect of any stray thermo emf in the system, we recorded
the voltage across the sample with the ends at zero temperature
difference ($V_{tare}$) in a different run under identical experimental
conditions. This stray voltage $V_{tare}$ was subtracted from
$V_{meas}$ at each temperature before calculating the absolute thermopower.
After the cooling run was over the heating data were collected at
every 1~K interval. The heating rate between two points was the
same as the cooling rate used earlier. This cycle of measurement
was also repeated with a different cooling and heating rate of
0.2~K/min.

It was observed that the thermopower above 200~K
does not show any time dependence, and is independent of
the history of the measurement. So to avoid the effect of any
previous measurements all the time dependent measurements used the
following protocol: for cooling cycle, first take the sample to
220~K, wait for half an hour, and then cool at 1.0~K/min to the
temperature of interest and once the temperature is stabilized
record the thermopower as a function of time. In the heating cycle
the time dependent thermopower was done in a similar way: first
take the sample to 220~K, wait for half an hour, then cool at 1.0~K/min
to 85~K and then heat at 1.0~K/min to the temperature of
interest, and once the temperature stabilized record the
thermopower as a function of time. It was ensured that the
temperature difference between the two ends of the sample was kept fixed at
2~K during all these measurements.

\begin{figure}[htp]
\begin{centering}
    \subfigure[]
    {\includegraphics[width=1\columnwidth]{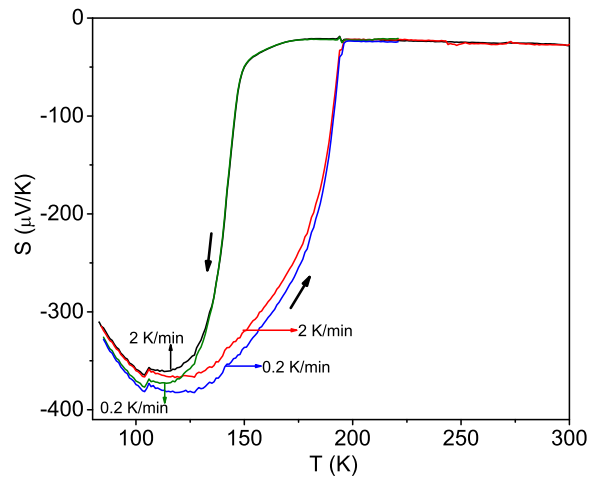}\label{fig:subfig1a}}
    \subfigure[]{\includegraphics[width=1\columnwidth]{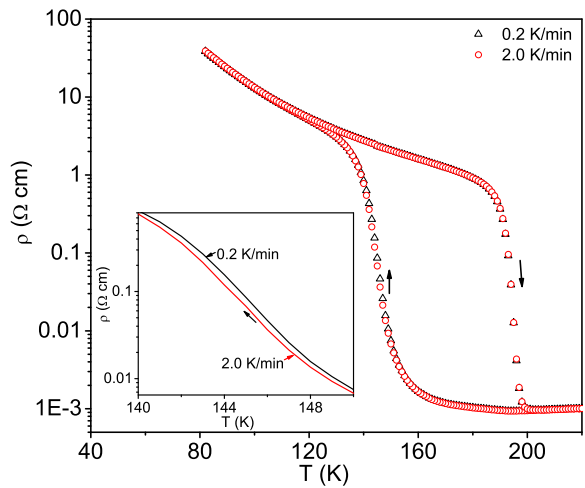}\label{fig:subfig1b}} \\
\end{centering}
\caption{(Color Online) (a) Temperature versus thermopower for
NdNiO$_3$ for the cooling/heating rate of 2.0~K/min and 0.2~K/min.
(b)Temperature versus resistivity for NdNiO$_3$ for the
cooling/heating rate of 2.0~K/min and 0.2~K/min. Inset shows the
expanded view of the resistivity plot, obtained while cooling,
close to the temperature where the metallic percolation ceases
(144 K).}
 \label{fig:S vs T}
\end{figure}

\section{Results}
 Figure \ref{fig:subfig1a} shows the thermopower of NdNiO$_{3}$ as a function of temperature.
 The thermopower is multiple valued, the cooling and heating data differing significantly from
 each other and forming a large hysteresis loop. The sign of thermopower remains negative both in the metallic and insulating state which indicates that the
 charge carriers are electrons in both states. On cooling the sample below 300~K, the magnitude of
 thermopower decreases in magnitude till 200~K.  Below this temperature the  thermopower shows a strong temperature
 dependence with its magnitude attaining a maximum at about 115~K. The thermopower data indicates that NdNiO$_3$
 undergoes a metal to insulator transition at 200~K which is consistent with the resistivity (see Figure \ref{fig:subfig1b} )
 and earlier thermopower measurements.\cite{Granados}  Below 115 K, the thermopower data does not follow
 the band gap model $(|S| \propto 1/T)$, unlike resistivity. A small jump in the thermopower has been observed at around  105~K, which is the region where the
 volume fraction of the metallic phases is small.  Such behavior has been predicted from Bergman and Levy's theory of effective thermopower of mixed phase systems
 and has been experimentally observed in Al-Ge thin films
 \cite{Bergman,Hurvits}.

\begin{figure}[!t]
\begin{centering}
\includegraphics[width=1\columnwidth]{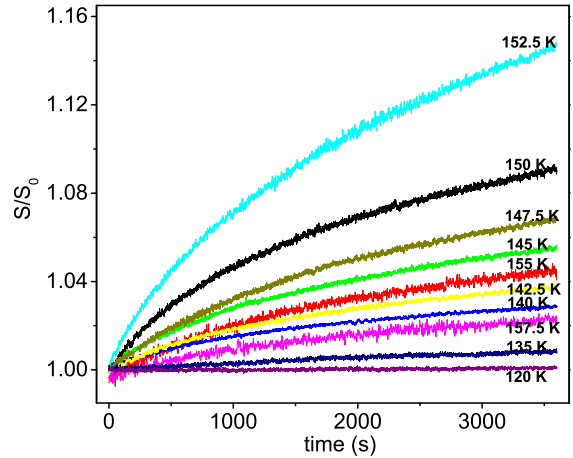}
\par\end{centering}
\caption{(Color Online) Time dependence of thermopower while cooling, at various temperatures in the range of 157.5-120 K, for a period of one hour}
\label{fig: S vs time}
\end{figure}

Below the M-I transition,  thermopower shows a significant
dependence on the rate of temperature change. See Figure
\ref{fig:subfig1a}. A slower cooling rate of 0.2~K/min yields a
higher absolute thermopower than that of the faster cooling rate
2.0~K/min. This rate dependence of thermopower is more readily
seen below 125~K during a cooling run, while in resistivity, the
cooling rate dependence is more significant near 145~K where the
percolation of the metallic regions cease. In the heating cycle
thermopower exhibits a large rate dependence at low temperatures
and this rate dependence persists upto about 180~K before
disappearing as we increase the temperature towards $T_{MI}$. In
contrast to this we did not find any significant heating rate
dependence in the  resistivity data(See Figure
\ref{fig:subfig1b}).The presence of cooling and heating rate
dependence in the thermopower data indicates that the system is
not in its thermodynamic equilibrium both in cooling and as well
as heating cycle. These observations are corroborated by the data
shown in Figure \ref{fig: S vs time} and Figure \ref{fig: S vs
Htime}.

 Figure \ref{fig: S vs time} displays a subset of the time dependent thermopower data recorded at fixed temperatures
 in the cooling run. The data are presented as $S(t)/S_0$, where $S_0 \equiv S(t=0)$, so that the values are normalized for easy comparison.
 We found that below 170~K, the absolute value of thermopower increases with time. A maximum relative increase in
 thermopower of about 14~\% was observed for a period of one hour at 152.5~K.  The time dependence of
 thermopower is lower both above and below this temperature. Below 120~K, we did not find any detectable
 time dependence in the system. All the time dependence data were fitted to a stretched exponential form
 \begin{equation}
S(t)=S_{0}+S_{1}\left(1-e^{-\left(\frac{t}{\tau}\right)^{\gamma}}\right)\label{eq:Stretched-Exponential}
\end{equation}
where $S_0$, $S_1$, $\tau$ and $\gamma$ are fit parameters with
 $S(t)$, $S_0$ and $S_1$ having negative value. The fits are
quite good with $R^2$ value greater than 0.995 in most cases. See
table 1.

\begin{figure}[!t]
\begin{centering}
\includegraphics[width=1\columnwidth]{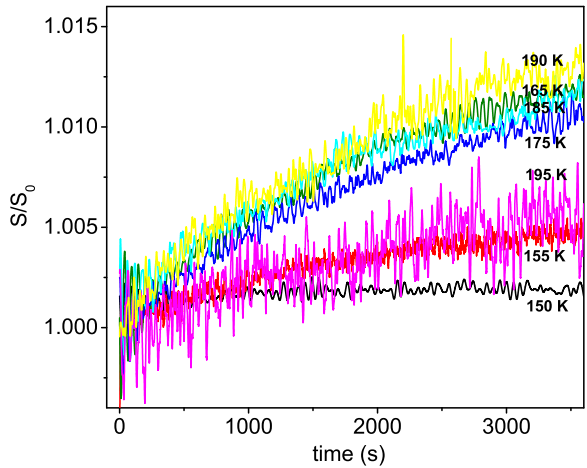}
\par\end{centering}
\caption{(Color Online) Time dependence of thermopower while
heating from 85 K, at various temperatures in the range of 150
-200 K, for a period of one hour}
\label{fig: S vs Htime}
\end{figure}

Figure \ref{fig: S vs Htime} shows a part of time dependent thermopower data recorded at fixed temperatures in the
 heating runs. We did not find any significant time dependence up to 150~K. Above 150~K, the magnitude
of thermopower increases with time. The maximum relative increase
in thermopower in one hour is around 1\% which  is much smaller
than that seen in the cooling runs. Above $T_{MI}$, the
thermopower data is stable and no time dependence or
cooling/heating rate dependence was observed .

\begin{table}
\begin{centering}
\begin{tabular}{|c|c|c|c|c|c|c|}
\hline
\# & T(K) & $S_{1}/S_{0}$ & $\tau$ ($10^{3}$s) & $\gamma$ & $\chi^{2}/DOF$ & $R^{2}$\tabularnewline
\hline
\hline
1 & 140.0 & 0.0426(1) & 3.12(2) & 0.729(1) & 0.34 & 0.99705\tabularnewline
\hline
2 & 145.0 & 0.0812(1) & 3.09(2) & 0.782(1) & 1.3 & 0.99837\tabularnewline
\hline
3 & 147.5 & 0.0905(2) & 2.53(1) & 0.896(1) & 2.3 & 0.99853\tabularnewline
\hline
4 & 150.0 & 0.118(1) & 2.34(1) & 0.826(1) & 3.2 & 0.99861\tabularnewline
\hline
5 & 152.5 & 0.231(1) & 3.62(2) & 0.776(1) & 5.3 & 0.99905\tabularnewline
\hline
6 & 155.0 & 0.0673(2) & 2.58(1) & 0.834(2) & 7.2$\times 10^3$ & 0.99687\tabularnewline
\hline
\end{tabular}
\par\end{centering}
\caption{\label{tab:Fit-parameters}Fit parameters for a subset of
time dependence data shown in Figure \ref{fig: S vs time}. The
degrees of freedom of the fits $DOF\approx1000$. The
$\chi^{2}/DOF$ for 155\,K is too large, indicating that we have
underestimated the error in thermopower in this case. Anyway, we
note that, the $R^{2}$ values are consistently good and indicate
reasonably good fits.}
\end{table}

\section{Discussion}The presence of out of equilibrium features in thermopower is similar to that observed
in our resistivity measurements reported earlier.\cite{Devendra}
Based on those observations we had proposed a model to understand
the behavior of the phase separated regime. The model is as
follows: The high temperature metallic phase survives below the
metal to insulator transition in its supercooled metastable state.
The supercooled metallic phase is not a single entity, but instead
is made up of various regions which can make a transition or
switch from the metallic to the insulating state independent of
the neighboring regions. These regions  are referred to as
switchable regions (SR). Each SR in its metallic state has to
cross  an energy barrier $U$ to go to its insulating state and the
barrier strength was determined to be $U=cV(T-T^*)^{1/4}$, where
$V$ is the volume of the SR, $T^*$ is its temperature of
metastability, and $c$ is a constant. An SR will switch to the
insulating state if (a) it  attains its temperature of
metastability or (b) it gains enough energy from thermal
fluctuations so that it can  cross the energy barrier $U$. For
$T^* <T <T_{MI}$ and $T_g<T$, supercooled phases will switch over
to the insulating  state stochastically  which will cause time
dependence in the physical properties. No time dependence in
resistivity  was observed in the heating cycle.\cite{Devendra}  On
the basis of this  it was claimed that all the SR's  switch to the
insulating state on cooling implying that  $T^* > T_g$ for all of
them.

 In thermopower measurements, during cooling, we got the maximum time dependence at around 152.5~K
while in the case of resistivity the maximum time dependence
occurs around 145~K. The maximum magnitude of change in
thermopower in one hour ($\Delta S$) is $\approx$ 14~\%, which is
much  smaller than that of resistivity ($\Delta \rho \approx
200\%$).\cite{Devendra}  We notice that the change in thermopower
in one hour $\Delta S$ is much closer to the change in insulating
volume fraction ($\Delta V \approx 8\%$).\cite{Devendra} The time
dependence of thermopower in the cooling cycle can be understood
on the basis of our model. As we have discussed in the previous
paragraph, below $T_{MI}$, the SR's present in the SC state
stochastically switch to the insulating state which enhances the
volume fraction of insulating phase with time. This increase in
insulating volume fraction results in a higher magnitude of
thermopower. Thermopower being a transport property, the change in
thermopower is also affected by percolation effects and thus the
behavior of time dependence of thermopower and the time dependence
of insulating volume fraction will not have a one-to-one monotonic
correspondence.

  Below 120~K no detectable time dependence was observed in the thermopower, but cooling
rates of 2~K/min and 0.2~K/min lead to slightly different low
temperature states with differing thermopower. This suggests that
a tiny fraction of SR's must remain in the supercooled state even
below 120~K. Since we did not find any time dependence below
120~K, we argue that the kinetics of these SR's are arrested and
and they end up in a glass-like state. Now, the question arises
why there is no rate dependence in resistivity below 120~K neither
in cooling runs nor in the subsequent heating runs. To answer this
question we suggest that the small fraction of SR's trapped in the
glassy state are not detectable in a resistivity measurement. This
idea is in consonance with the fact that thermopower measurements
show  a small time dependence of about 1\% over a period of an
hour in heating runs (Figure \ref{fig: S vs Htime}) while no
detectable time dependence  was observed in the heating runs of
resistivity measurements.\cite{Devendra}

The advantage of thermopower over resistivity in detecting  tiny
amounts of SC metallic regions (SR's) dispersed in an  insulating
matrix can perhaps be understood as follows. When the fraction of
SC metallic regions is small they will be completely surrounded by
insulating regions. Now the net contribution of a tiny metallic
region embedded in an insulating matrix on the total resistivity
of the material will be determined by the resistance of the
metallic region plus its inter-crystallite resistances in the
current path. Now we conjecture, based on our experience on the
contact resistance of polycrystalline NdNiO$_3$, that the
inter-crystallite resistance between a metallic and an insulating
crystallite will be very high compared to the resistance of the
metallic crystallite. This implies that the tiny metallic region
will be essentially shut out of the current path and will not be
able to affect the resistivity of the material significantly. The
boundary effects are much less significant in the case of
thermopower compared to resistivity because the thermal
conductivities of the metallic and insulating regions are
comparable and this makes the thermopower sensitive to the
embedded metallic regions.\cite{Cooper,Jhou}

 In a heating run, the SR's present in a glassy state
will remain arrested in that state till their respective $T_g$'s
are attained. In cooling runs we found time dependence down to
120~K while in heating runs we did not find any time dependence
until 150~K which suggests that the SR's which are arrested in the
glassy state have their $T_g$'s above 150~K.  On heating the
system above 150~K, the kinetics of the SR's gets dearrested and
it can cross the free energy barrier $U$ and flip to the
insulating state with  a probability of flipping  $\propto
exp(-U/k_BT)$. In the time dependence data presented in Figure
\ref{fig: S vs Htime} the flipping of the SR's to the insulating
state raises the insulating volume fraction with time and this
enhances the magnitude of thermopower with time. The fact that we
get noticeable time dependence all the way upto $T_{MI}$ suggests
that the $T_g$'s are distributed quite broadly upto $T_{MI}$. We
might now say that each SR has a unique temperature of
metastability $T^*$ and a unique temperature of kinetic arrest
$T_g$. In a cooling run, the majority of the SR's present in the
supercooled state switch to the insulating state by the time their
$T^*$ is attained and only a small fraction of SR's get arrested
in the glassy state and this suggests that the majority of the
SR's have their $T^*$ greater than $T_g$.

\section{Conclusion}
Our experimental results and analyses suggest the following
picture for the NdNiO$_3$ system. Below $T_{MI}$  the system
consists of SC metallic and insulating regions. Each of the SC
metallic region  has a unique temperature of metastability $T^*$
and a unique temperature of kinetic arrest $T_g$. For the majority
of the SC regions  $T^*$ is greater than $T_g$. Thus when the
sample is cooled below $T_{MI}$, while the majority of the SC
regions switch to insulating state when their temperature of
metastability is reached, a small fraction of the SC regions which
have $T_g > T^*$ get arrested in a glassy state and remain in that
state down to low temperature. Such SC regions have their $T_g$
greater than about 150~K. In a heating run, these SC metallic
regions remain trapped in their glassy metallic state till their
temperature of kinetic dearrest ($T_g$) is reached. Above this
temperature, these regions switch over to the insulating state
stochastically.

\section{Acknowledgement} DK thanks the University Grants Commission
of India for financial support. JAA and MJM-L acknowledge the
Spanish Ministry of Education for funding the Project
MAT2007-60536  .

\

\end{document}